\documentclass[aps,pra,showpacs,twocolumn,floatfix,superscriptaddress]{revtex4}

\usepackage{dcolumn,bm,graphicx,amsmath,amsfonts,amssymb}
\usepackage{hyperref}

\begin{document}



\date{\today}

\title{Precision determination of weak charge of $^{133}$Cs from atomic parity violation}

\newcommand{\Reno}{
Physics Department, University of Nevada,
Reno, Nevada  89557, USA}
\newcommand{\Sydney}{
School of Physics, University of New South Wales,
Sydney, NSW 2052, Australia}
\newcommand{\Gatchina}{
Petersburg Nuclear Physics Institute,
Gatchina, Leningrad district, 188300, Russia}
\newcommand{\Auckland}{
Centre for Theoretical Chemistry and Physics,
The New Zealand Institute for Advanced Study,
Massey University Auckland, Private Bag 102904, 0745, Auckland, New Zealand}

\author{S.~G.~Porsev}
\affiliation{\Reno}
\affiliation{\Sydney}
\affiliation{\Gatchina}

\author{K.~Beloy}
\affiliation{\Reno}
\affiliation{\Auckland}

\author{A.~Derevianko}
\affiliation{\Reno}

\begin{abstract}
We discuss results of the most accurate to-date test of the low-energy electroweak sector of the standard model of elementary particles. Combining previous measurements with our high-precision calculations we extracted the weak charge of the $^{133}$Cs nucleus,
$Q_W = -73.16(29)_\mathrm{exp}(20)_\mathrm{th}$ [S.~G.~Porsev, K.~Beloy, and A.~Derevianko, Phys.~Rev.~Lett.~{\bf 102}, 181601 (2009)].
The result is in perfect agreement with $Q_W^\mathrm{SM}$ predicted by the standard model, $Q_W^\mathrm{SM} =-73.16(3)$, and confirms energy-dependence (or running) of the electroweak interaction and places constraints on a variety of new physics scenarios beyond the standard model. In particular, we increase the lower limit on the masses of extra $Z$-bosons predicted by models of grand unification and string theories.  This paper provides additional details to the Letter. We discuss large-scale calculations in the framework of the coupled-cluster method, including full treatment of single, double, and valence triple excitations. To determine the accuracy of the calculations we computed energies, electric-dipole amplitudes, and hyperfine-structure constants. An extensive comparison with high-accuracy experimental data was carried out.

\end{abstract}


\maketitle


\section{Introduction}
Atomic parity violation (APV) places powerful constraints on
new physics beyond the standard model of elementary particles~\cite{MarRos90}.
The APV measurements in
are interpreted in terms of the weak nuclear
charge $Q_W$, quantifying the strength of the electroweak coupling between atomic
electrons and quarks of the nucleus. At the tree level the weak nuclear charge is given
by a simple formula
\begin{equation}
Q_W = -N + Z\,(1-4\,\sin^2\theta_W),
\end{equation}
where $N$ is the number of neutrons, $Z$ is the nuclear charge,
and $\theta_W$ is the Weinberg angle.
Since $\sin^2\theta_W$ is close to 0.23, the weak nuclear charge $Q_W$ is numerically
close to $-N$.

In Ref.~\cite{PorBelDer09} we reported the most accurate to-date
determination of this coupling strength by combining previous measurements~\cite{WooBenCho97,BenWie99} with
high-precision calculations in cesium atom. We found the result
$Q_W(^{133}\mathrm{Cs}) = -73.16(29)_\mathrm{exp}(20)_\mathrm{th}$~\cite{PorBelDer09} to be in a perfect agreement
with $Q_W^\mathrm{SM}$ predicted by the standard model (SM), $Q_W^\mathrm{SM} =-73.16(3)$~\cite{Ams08}.
In this work we provide a detailed account of the calculation carried out in Ref.~\cite{PorBelDer09}.


Historically, APV helped to establish the validity of the SM~\cite{Khr91,BouBou97,GinFla04}.
While a  number of APV experiments have been carried out~\cite{EdwPhiBai95,VetMeeMaj95,MeeVetMaj95,PhiEdwBai96,MacZetWar91,TsiDouSta09,TsiDouSta10},
the most accurate  measurement is due to Wieman and collaborators~\cite{WooBenCho97}.
They determined the ratio of $E_\mathrm{PNC}/\beta=1.5935(56) \, \mathrm{mV/cm}$
(where $E_\mathrm{PNC}$ is the parity nonconserving amplitude defined below by Eq.~(\ref{Eq:EPNC})
and $\beta$ is the vector transition polarizability) on the forbidden $6S_{1/2} \rightarrow 7S_{1/2}$
transition in atomic Cs with an accuracy of 0.35\%. This measurement does not directly translate into an
electroweak observable of the same accuracy, as
the interpretation of the experiment requires input from atomic theory, which links  $Q_W$ to the signal.
$Q_W$ is treated as a parameter and by combining $E_\mathrm{PNC}$ calculations  with
measurements, the value of $Q_W$ is extracted and can be compared with the SM value either
revealing or constraining new physics.

The parity nonconserving (PNC) amplitude for the $6S_{1/2} \rightarrow 7S_{1/2}$ transition
in Cs may be evaluated as
\begin{eqnarray}
\lefteqn{E_\mathrm{PNC} = \sum_{n}
\frac{\langle 7S_{1/2}|D_0|nP_{1/2}\rangle  \langle nP_{1/2} |H_{W}|6S_{1/2}\rangle
}{E_{6S_{1/2}}-E_{nP_{1/2}}}  } \nonumber \\  &+ &
\sum_{n}
\frac{\langle 7S_{1/2}|H_{W}|nP_{1/2}\rangle  \langle nP_{1/2} |D_0|6S_{1/2}\rangle
}{E_{7S_{1/2}}-E_{nP_{1/2}}}
\, .
\label{Eq:EPNC}
\end{eqnarray}
Here $D$ and $H_W$ are electric-dipole and weak interaction operators,
and $E_{i}$ appearing in the denominators are atomic energy levels. The effective weak interaction mediated by
$Z$-bosons averaged over quarks reads
$
 H_W = -\frac{G_F}{\sqrt{8}} \, Q_W \,  \gamma_5 \,
 \rho ({\bf r}) $,
where $G_F$ is the Fermi constant, $\gamma_5$ is the Dirac matrix, and
$\rho ({\bf r})$ is the  neutron-density distribution.

\begin{figure}[h]
\begin{center}
\includegraphics*[scale=0.35]{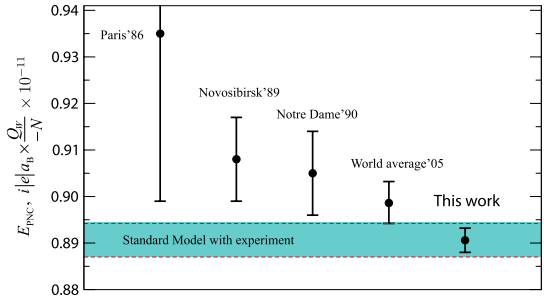}
\caption{(Color online)  Progress in evaluating the PNC amplitude.
Points marked Paris '86, Novosibirsk '89, Notre Dame '90
correspond to Refs.~\cite{BouPik86}, \cite{DzuFlaSus89}, and~\cite{BluJohSap90}.
Point ``World average '05'' is due to efforts of several groups~\cite{Der00,Der02,theorPNCextraNewBefore2005,MilSusTer02,ShaPacTup05}
on sub-1\% Breit, QED, and neutron-skin corrections reviewed in
Ref.~\cite{DerPor07}. The strip corresponds to a combination of the Standard Model $Q_W$ with
measurements~\cite{WooBenCho97,BenWie99}. The edges of the strip correspond to $\pm \sigma$ of the measurement.
Here we express $E_\mathrm{PNC}$ in
conventional units of $ i |e| a_{\mathrm B} \left( -{Q_W}/{N} \right) \times 10^{-11}$,
where $e$ is the elementary charge and $a_{\mathrm B}$ is the Bohr radius. These units
factor out a ratio of $Q_W$ to its approximate value, $-N$.
\label{Fig:compEPNC}}
\end{center}
\end{figure}

Interpretation of the PNC measurements requires evaluating Eq.~(\ref{Eq:EPNC}).
Although the underlying theory of quantum electrodynamics (QED) is well established, the atomic many-body problem is intractable.
Reaching theoretical accuracy
equal to or better than the experimental accuracy of 0.35\%
has been a challenging task (see Fig.~\ref{Fig:compEPNC}).
An important  1\% accuracy  milestone has been reached by the Novosibirsk~\cite{DzuFlaSus89}
and Notre Dame~\cite{BluJohSap90} groups in the late 1980s.
More recently, several groups have contributed to understanding sub-1\% corrections, primarily due to
the Breit (magnetic) interaction and radiative QED processes~\cite{Der00,Der02,theorPNCextraNewBefore2005,MilSusTer02,ShaPacTup05} (reviewed in \cite{DerPor07}). The results of these calculations
are summarized by the ``World average '05'' point of Fig.~\ref{Fig:compEPNC}, which has a 0.5\% error bar reflecting this progress.
As of 2005, the sensitivity to new physics has been limited by the accuracy of solving the basic correlation problem.
Here we report an important progress in solving it.

We wish to  evaluate accurately the sum~(\ref{Eq:EPNC}).
Cs atom has one loosely-bound valence electron $v=6s_{1/2}, 6p_{1/2}, ...$ outside a closed-shell core
$1s^2\, 2s^2\, 2p_{1/2}^2 \cdots 5d_{5/2}^6$. We compute atomic wave functions, energies, and
matrix elements and sum over the intermediate states. We solve the eigenvalue problem $H |\Psi_v\rangle = E_v |\Psi_v\rangle$
and find atomic wave functions and energies.
Our specific scheme~\cite{DerPor05,PorDer06Na,DerPor07,DerPorBel08} of solving the atomic
many-body problem is rooted in the coupled-cluster method~\cite{LinMor86}. Details will be provided
in the next section.

The paper is organized as follows.
In Sec.~\ref{sec_CCSDvT} we describe the coupled-cluster (CC) approximation
including single, double, and valence triple excitations and present the
results for the low-lying energy levels.
In Sec.~\ref{sec_MBPT} we evaluate the hyperfine structure constants, the matrix
elements of the electric dipole moment, and the PNC amplitude. An analysis of
uncertainty of the PNC amplitude is also presented.
In Sec.~\ref{sec_Qw} we extract the weak nuclear charge from
the theoretical and experimental quantities and discuss implications
for particle physics.
If not stated otherwise the atomic units ($\hbar = |e| = m_e = 1$)
are used throughout.

\section{Coupled-cluster approximation}
\label{sec_CCSDvT}
We employ an approximation
rooted in the coupled-cluster method~\cite{CoeKum60,Ciz66}. The
key difference compared to the previous CC-type calculations for
univalent atoms (see, e.g.,
Refs.~\cite{BluJohLiu89,BluJohSap91,EliKalIsh94,AvgBec98,SafDerJoh98,SafJohDer99})
is our additional inclusion of valence triple excitations in the
expansion of the cluster amplitude. We refer to this approximation
as the coupled-cluster single, double, and valence triple (CCSDvT)
method. Details of our approximation may be found in
Refs.~\cite{DerPor05,PorDer06Na,DerPor07,DerPorBel08}. Below we
briefly recapitulate its main features and present numerical results.

We  choose the  lowest-order Hamiltonian to include the relativistic kinetic
energy operator of electrons and their interaction with the nucleus and the
$V^{N-1}$ (frozen-core) Dirac-Hartree-Fock (DHF) potential.
The single-particle orbitals and energies  $\varepsilon_i$ are found from the set of
the frozen-core DHF equations.
 With the DHF single-particle orbitals, the second-quantized Hamiltonian  reads
\begin{eqnarray}
 H &=&  H_0 + G \nonumber \\
   &=& \sum_{i} \varepsilon_i N[a_i^\dagger a_i] +
 \frac{1}{2} \sum_{ijkl} g_{ijkl} N[a^\dagger_i a^\dagger_j a_l a_k ] \, .
\label{Eq:SecQuantH}
\end{eqnarray}
 Here $H_0$ is the one-electron lowest-order Hamiltonian,
$G$ is the residual Coulomb interaction,
$a_{i}^\dagger$ and $a_{i}$ are the creation and
annihilation operators, and $N[\cdots]$ is the normal product
of operators with respect to the core quasi-vacuum state
$|0_c\rangle$. Indices $i, j,k$, and $l$ range over all possible
single-particle orbitals, and
$g_{ijkl}$ are the  Coulomb matrix elements.

The exact many-body state $|\Psi_v\rangle$ can be represented as follows
\begin{eqnarray}
|\Psi_v\rangle &=& N[ \exp(K) ]\, |\Psi_v^{(0)}\rangle \nonumber \\
&=& \left( 1 + K + \frac{1}{2!} N[K^2] + \ldots \right)
\, |\Psi_v^{(0)}\rangle \, ,
\label{Eq:PsivOmega}
\end{eqnarray}
where $|\Psi_v^{(0)} \rangle$ is the lowest-order DHF state.
The cluster operator $K$ is expressed in terms of connected
diagrams of the wave operator.
In our approach the operator $K$ is approximated by
\begin{eqnarray}
&& K \approx S_c + D_c + S_v + D_v +T_v = \label{Eq:KCCSDvT} \nonumber \\
&&\begin{array}{l}
\includegraphics*[scale=0.4]{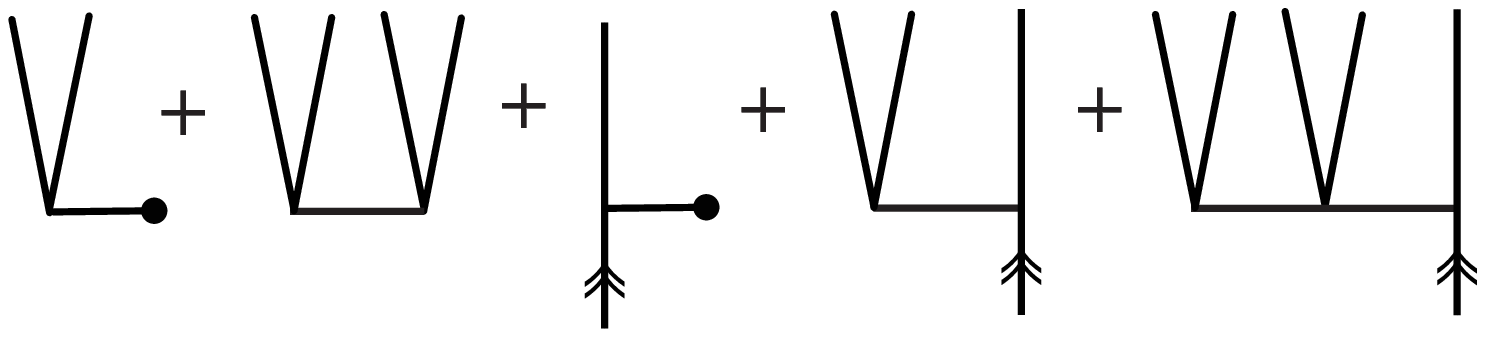}
\, ,
\end{array}
\end{eqnarray}
with the double-headed arrow representing the valence state.
Here $S_v$ and $D_v$ ($S_c$ and $D_c$)
are the valence (core) singles and doubles, and $T_v$ are the valence triples.
Compared to the core excitations, the valence amplitudes involve excitations of the valence electron.
A popular singles-doubles (SD) approximation~\cite{BluJohLiu89} corresponds to neglecting non-linear terms in the expansion~(\ref{Eq:PsivOmega}) and valence triples in Eq.~(\ref{Eq:KCCSDvT}).

Below we present the topological structure of the equations for valence cluster amplitudes
in the CCSDvT approximation. The equations in the coupled-cluster singles-doubles approximation (i.e., without triples) are presented in
explicit form in Ref.~\cite{PalSafJoh07}.  A detailed tabulation of the formulas for
the valence triple amplitudes is given in Ref.~\cite{PorDer06Na}.
Since we do not take core triples into consideration, the CCSDvT equations for the core
amplitudes $S_c$ and $D_c$ are the same as in~\cite{PalSafJoh07}.

The total energy of the valence electron is given by the sum of the DHF value and the correlation energy, $\delta E_v$,
\begin{equation}
E^{\rm tot}_{\rm CCSDvT} = E_{\rm DHF} + \delta E_v
\label{Eq:E_tot} \, .
\end{equation}
Following the notation of Ref.~\cite{DerPorBel08},
we can represent the correlation valence energy $\delta E_v$ as
\begin{equation}
\delta E_v = \delta E_\mathrm{SD} + \delta E_\mathrm{CC} + \delta
E_\mathrm{vT} \label{Eq:delta_Ev} \, ,
\end{equation}
where correction $\delta E_\mathrm{SD}$ is obtained within
the SD approach, correction $\delta E_\mathrm{CC}$ comes from
nonlinear CC contributions, and $\delta E_\mathrm{vT}$ is due to
valence triples.

The topological structure for the valence singles and valence doubles equations may be
represented as~\cite{DerPorBel08}
\begin{eqnarray}
&&- [H_0, S_v]  + \delta E_{v} S_v \approx  {\rm SD} + \nonumber \\
&&  S_v[S_c \otimes S_v]  + S_v[S_c \otimes S_c] + \nonumber \\
&&  S_v[S_c \otimes D_v]  +  S_v[S_v \otimes D_c] + S_v[T_v]  \, .
\end{eqnarray}
\begin{eqnarray}
&&- [H_0, D_v]  + \delta E_{v} D_v \approx  {\rm SD} +\nonumber \\
&&  D_v[S_c \otimes S_v] + D_v[S_c \otimes S_c]+ \nonumber \\
&&  D_v[S_c \otimes D_v] +D_v[S_v \otimes D_c] + D_v[S_c \otimes D_c] + \nonumber\\
&&  D_v[D_c \otimes D_v] + D_v[T_v] \, .
\label{Eq:SvDv}
\end{eqnarray}
Here $[H_0, S(D)_v]$ are commutators. $S_v[S_c \otimes S_v]$ stands for a contribution resulting from  a
product of clusters $S_c$ and $S_v$. All other
terms are defined in a similar fashion. SD terms encapsulate contributions from the SD approximation~\cite{BluJohLiu89}.

For valence triple amplitudes we obtain symbolically,
\begin{equation}
- [H_0, T_v]  + \delta E_{v} T_v
 \approx
 T_v[D_c]+ T_v[D_v]  \,.
\end{equation}
Contributions $T_v[D_c]$ and $T_v[D_v]$ denote the effect of core
and valence doubles on valence triples, respectively.
In the present analysis we include only these effects, while
omitting the effect of valence triples on valence triples
and nonlinear CC contributions~\cite{DerPorBel08}.
These are higher-order effects that are prohibitively
time-consuming for the 55-electron Cs atom.

A numerical solution of the CCSDvT equations provides us with the cluster amplitudes and
correlation energies. With the obtained wave functions for two valence states $w$ and $v$
we may evaluate various matrix elements (MEs),
\begin{equation}
 Z_{wv} =
 \frac{ \langle \Psi_w | \sum_{ij} \langle i | z | j\rangle \, a^\dagger_i a_j | \Psi_v \rangle }
 { \sqrt{ \langle \Psi_w | \Psi_w \rangle \langle \Psi_v | \Psi_v \rangle  }}
 \label{Eq:Zmel} \, .
\end{equation}
The corresponding CCSDvT expressions are given in Ref.~\cite{PorDer06Na}.  There are two important modifications compared
to the earlier computations~\cite{BluJohSap90}: (i) explicit inclusion of valence triples in the expressions for matrix
elements and (ii) dressing of lines and vertices in expressions for matrix elements.
The dressing mechanism~\cite{DerPor05} may be explained as follows: when the CC exponent
is expanded in Eq.~(\ref{Eq:Zmel}), we encounter an infinite number of terms.
The resulting series may be partially summed by considering the topological structure of
the product of
cluster amplitudes, which may be classified using the language of $n$-body insertions. We include
two types of insertions: particle-- and hole--line insertions (line
``dressing'') and two-particle and two-hole
random-phase-approximation-like  insertions.

Our  CCSDvT code is an extension of the relativistic SD
code~\cite{SafDerJoh98} which employs a B-spline basis
set~\cite{JohBluSap88}. Our present version uses a more robust
dual-kinetic-balance B-spline basis set~\cite{ShaTupYer04} as
described in Ref.~\cite{BelDer08}. This basis  numerically
approximates a complete set  of single-particle atomic orbitals.
Here, for each partial wave $\ell$ we use 35 out of $N_\mathrm{bas}
=40$ positive-energy  basis functions generated in a cavity of
radius $R_\mathrm{cav}=75$ bohr. Basis functions with $\ell_{\rm
max} \le 5$ are used for single and double excitations. For triple
excitations we employ a more limited set of basis functions with
$\ell_{\rm max}(T_v) \le 4$. Excitations from core sub-shells
[4$s$,...,5$p$] are included in the calculations of triples while
excitations from sub-shells [1$s$,...,3$d$] are discarded. A basis
set extrapolation correction to  infinitely large $\ell_{\rm max}$,
$N_\mathrm{bas}$, and $R_\mathrm{cav}$ is added separately.

Computations were done on a non-uniform grid of 500 points with 15
points inside the nucleus. The nuclear charge distribution was
approximated by  $\rho(r) = \rho_0/(1+\exp[(r-c)/a])$ both when
solving the DHF equations and evaluating weak interaction matrix
elements. For $^{133}$Cs, $c = 5.6748$ fm and $a=0.52338 $  fm.


 Numerical results for the energies are presented
in Table~\ref{Tab:Cs_E}. The dominant
contribution to the energies comes from the DHF values.
Correlation corrections ($\delta E_\mathrm{SD}$,
$\delta E_\mathrm{CC}$, and $\delta E_\mathrm{vT}$) are dominated by the SD contribution.
We also incorporate small complementary corrections due to the
Breit interaction, basis extrapolation ($\delta E_{\rm extrap}$),
and quantum-electrodynamic (QED) radiative corrections. The
 agreement between our {\em ab initio}  and experimental values is at the level
of 0.3\% for the $6S$ state and 0.1-0.2\% for all other states.
\begin{table}[h]
\caption{Contributions to removal energies of $6S_{1/2}$,
$6P_{1/2}$, $7S_{1/2}$, and $7P_{1/2}$ states for Cs in cm$^{-1}$ in
different approximations. A comparison with experimental values is
presented in the bottom panel. } \label{Tab:Cs_E}
\begin{ruledtabular}
\begin{center}
\begin{tabular}{lrrrr}
\smallskip
& \multicolumn{1}{c}{$6S_{1/2}$}
& \multicolumn{1}{c}{$6P_{1/2}$}
& \multicolumn{1}{c}{$7S_{1/2}$}
& \multicolumn{1}{c}{$7P_{1/2}$}\\
\hline
\smallskip
$E_{\rm DHF}$               &  27954   &  18790   &  12112   &  9223 \\
$\delta E_{\rm SD}$         &   3868   &   1610   &    827   &   460 \\
\smallskip
$\delta E_{\rm CC}$         & $-$379   & $-$178   &  $-$60   & $-$43 \\
\smallskip
$\delta E_{\rm vT}$         & $-$151   &  $-$44   &  $-$30   & $-$12 \\
\smallskip
$E^{\rm tot}_{\rm CCSDvT}$  &  31292   &  20178   &  12849   &  9628 \\
$\delta E_{\rm Breit}$\footnotemark[1]
                            &      2.6 &   $-$7.1 &      0.3 & $-$2.5 \\
$\delta E_{\rm QED}$\footnotemark[2]
                            &  $-$17.6 &   $-$4.1 &   $-$0.4 & $-$0.1 \\
$\delta E_{\rm extrap}$
                            &     32.2 &     15.5 &      7.1 &    4.6 \\
\hline
$E^{\rm tot}_{\rm final}$& 31309   &  20182   &  12856   & 9630  \\
{\rm $E_{\rm experim}$}\footnotemark[3]
                            &  31406   &  20228   & 12871    & 9641 \\
\end{tabular}
\end{center}
\end{ruledtabular}
\footnotemark[1]{Ref.~\cite{Der01Br}};
\footnotemark[2]{Ref.~\cite{FlaGin05}};
\footnotemark[3]{Ref.~\cite{Moo58}}.
\end{table}

Since the CCSDvT method is an approximation, we miss certain correlation effects
(due to omitted quadruple and higher-rank excitations). This
is the cause of the difference between computed and experimental energies in Table~\ref{Tab:Cs_E}.
To partially account for the missing contributions in calculations of matrix elements,
we additionally correct the CCSDvT wave functions using a semi-empirical procedure suggested in Ref.~\cite{BluJohSap92} (see justification in Ref.~\cite{DerPorBel08}).
In this approach, the valence singles, $S_v$, are rescaled by the ratio of
experimental and theoretical correlation energies.
A consistent definition of the experimental correlation energies ($\delta E^{\rm exp}_v$) requires removing the Breit and QED corrections from the experimental energy, i.e.,
\begin{equation}
\delta E^{\rm exp}_v = E_{\rm exp} - E_{\rm DHF} - \delta E_{\rm Breit} - \delta E_{\rm QED} .
\label{Eq:delE_exp}
\end{equation}
We will refer to  results obtained using the described procedure as ``scaling''.

\section{Evaluation of the parity nonconserving amplitude and
supporting quantities}
\label{sec_MBPT}

Below we present details for the evaluation of  the parity nonconserving amplitude
for the $6S_{1/2} \rightarrow 7S_{1/2}$ transition given by Eq.~(\ref{Eq:EPNC}).
The CCSDvT method is an approximation, and
an important part of the entire problem lies with evaluating the theoretical accuracy of the computed PNC amplitude. The PNC
amplitude cannot be directly compared to an experimental
measurement. As seen from Eq.~(\ref{Eq:EPNC}),
one needs to know the matrix elements of the electric-dipole operator,
the energies, and the matrix elements of the weak interaction $H_W$.
The quality of calculations of the dipole transition amplitudes and energy levels can be established
by comparing them with experimental data, while for the matrix
elements of the weak interaction such a direct  comparison is not possible. Instead,
 we may consider
the operator of the hyperfine interaction.
Matrix elements of both the hyperfine and weak interaction are accumulated near the origin.
Therefore, calculating hyperfine structure (HFS) constants for the low-lying states and comparing
them with the experimental data allows us to assess the quality of the constructed
wave functions near the nucleus.

\begin{figure}[h]
\begin{center}
\includegraphics*[scale=0.3]{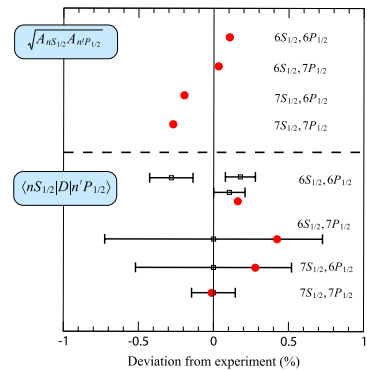}
\caption{(Color online)  Deviations of  computed values (red filled circles) from experimental data (centered at zero).
The upper panel displays combinations of magnetic hyperfine structure constants $\sqrt{A_{nS_{1/2}} A_{n'P_{1/2}}}$
which mimic matrix elements of the weak interaction. For these combinations, experimental error bars are
negligible compared to the theoretical accuracy.
The lower panel exhibits deviations of the computed dipole matrix elements from the most accurate experimental results~\cite{lifetimesCsRefs,VasSavSaf02}.
\label{Fig:comp}}
\end{center}
\end{figure}

The results of calculations of the HFS constants and dipole matrix
elements between the low-lying states are presented in
Tables~\ref{Tab:Ahfs} and~\ref{Tab:E1_ME}. For the HFS calculations
we assumed a uniform distribution of the nuclear magnetization
(magnetization radius of 5.6748  fm) and used the nuclear $g$-factor
of 0.73772. We explicitly list DHF and SD values. The entry
$\Delta$(CC) indicates the change in the value caused by including
the nonlinear terms in the equations for core and valence singles
and doubles. Likewise, $\Delta$(vT) and  $\Delta$(scaling) arise due
to a subsequent addition of valence triples and scaling. We also
incorporated smaller corrections: line and vertex dressing
(discussed in detail in~\cite{DerPor05}), the Breit interaction, the
QED corrections, and the corrections due to the basis set
extrapolation. In the lower panels of the tables we compare our
theoretical results with the most accurate experimental results. We
find that the discrepancies between theoretical and experimental
values for the HFS constants are 0.15-0.35\%.  For dipole matrix
elements the theoretical values are within the error bars of the
experiments.
The uncertainty estimate of $E_\mathrm{PNC}$ can be carried out using
geometric means $\sqrt{A_{nS_{1/2}} A_{n'P_{1/2}}}$~\cite{DzuFlaGin02}, as
the relative uncertainty of this combination mimics the relative uncertainty of
the ME of the weak interaction $\langle nS_{1/2} |H_W| n'P_{1/2} \rangle$.
Deviations of these combinations from experimental data are shown in the
upper panel of Fig.~\ref{Fig:comp}.
We find that the standard deviation of our theoretical values for these combinations from the experimental values is 0.2\%.
\begin{table*}[h]
\caption{Magnetic-dipole hyperfine structure constants $A$ (in MHz)
for $^{133}$Cs. Results of calculations and comparison with experimental
values are presented. See text for the explanation of entries.}
\label{Tab:Ahfs}
\begin{ruledtabular}
\begin{tabular}{ldddd}
& \multicolumn{1}{c}{$A(6S_{1/2})$}   & \multicolumn{1}{c}{$A(6P_{1/2})$}
& \multicolumn{1}{c}{$A(7S_{1/2})$}   & \multicolumn{1}{c}{$A(7P_{1/2})$} \\
\hline
DHF                            &  1424.9  &  160.90  &   391.53   &  57.61\\
\smallskip
SD                             &  2436.7  &  310.73  &   560.85   &  98.34\\
\smallskip
$\Delta$(CC)                   &  -110.1  &  -22.40  &   -13.43   &  -5.26 \\
\smallskip
$\Delta$(vT)                   &   -29.6  &    2.35  &    -1.92   &   1.18 \\
$\Delta$(scaling)              &    14.7  &    2.59  &     0.38   &   0.41 \\
{\it Complementary corrections}: \\
Line \& vertex dressing       &    -9.4  &   -1.92  &    -1.22   &  -0.53  \\
\smallskip
Breit\footnotemark[1]          &     4.9  &   -0.52  &     1.15   &  -0.15  \\
\smallskip
QED                            &    -9.7\footnotemark[2]
                                          &   -0.05\footnotemark[3]
                                                     &   -2.30\footnotemark[4]
                                                                  & -0.02\footnotemark[4]   \\
\smallskip
Basis extrapolation      &     9.1  &    0.71  &     1.08   &  0.10   \\
\smallskip
Final results                  &  2306.6  &  291.49  &   544.59   & 94.07    \\
Experiment                     &  2298.16 &  291.9135(15)\footnotemark[5]
                                                  &  545.90(9)\footnotemark[6]
                                                               & 94.35(4)\footnotemark[7] \\
                               &          &  291.9309(12)\footnotemark[8] &&  \\
Difference                    &   0.36\% &  -0.15\%  &    -0.24\%  &  -0.30\%  \\
\end{tabular}
\end{ruledtabular}
\footnotemark[1]{Ref.~\cite{Der01Br}};
\footnotemark[2]{Ref.~\cite{SapChe03}};
\footnotemark[3]{Ref.~\cite{SapChe06}};
\footnotemark[4]{The QED corrections for the $7S_{1/2}$ and $7P_{1/2}$ states were obtained
by scaling those for the $6S_{1/2}$ and $6P_{1/2}$ states};
\footnotemark[5]{Ref.~\cite{DasNat08}};
\footnotemark[6]{Refs.~\cite{GilWatWie83}};
\footnotemark[7]{Ref.~\cite{FeiSahPut72}};
\footnotemark[8]{Ref.~\cite{GerCalTan06}}.
\end{table*}
\begin{table*}[h]
\caption{ Reduced matrix elements of the electric dipole moment operator $D$ (in a.u.) for
$^{133}$Cs. Results of calculations and comparisons with experimental
values are presented. See text for the explanation of entries.
\label{Tab:E1_ME} }
\begin{ruledtabular}
\begin{tabular}{ldddd}
& \multicolumn{1}{c}{$|\langle 6P_{1/2} ||D|| 6S_{1/2} \rangle|$} &
\multicolumn{1}{c}{$|\langle 7P_{1/2} ||D|| 6S_{1/2} \rangle|$} &
\multicolumn{1}{c}{$|\langle 6P_{1/2} ||D|| 7S_{1/2} \rangle|$}
& \multicolumn{1}{c}{$|\langle 7P_{1/2} ||D|| 7S_{1/2} \rangle|$} \\
\hline
DHF                             &  5.2777     &  0.3717  &   4.4131   &  11.009\\
\smallskip
SD                              &  4.4831     &  0.2969  &   4.1984   &  10.256\\
\smallskip
$\Delta$(CC)                    &  0.0717     &  0.0058  &   0.0528   &   0.045\\
\smallskip
$\Delta$(vT)                    & -0.0423     & -0.0302  &   0.0038   &   0.009\\
$\Delta$(scaling)               & -0.0123     &  0.0033  &  -0.0138   &  -0.010\\
\smallskip
{\it Complementary corrections}: \\
\smallskip
Line \& vertex dressing         &  0.0036     &  0.0016  &  -0.0004   &   0.001 \\
\smallskip
Breit\footnotemark[1]          & -0.0010     &  0.0019  &   0.0049   &  -0.003 \\
\smallskip
QED                             &  0.0027\footnotemark[2]\footnotemark[3]
                                              & -0.0028\footnotemark[3]
                                                         &  -0.0043\footnotemark[3]
                                                                      &   0.005\footnotemark[3]  \\
\smallskip
Basis extrapolation                 &  0.0038     &  0.0005  &   0.0036   &   0.005  \\
\hline
Final result                    &  4.5093     &  0.2769  &   4.2450   &  10.307  \\
Experiment                      &  4.5097(45)\footnotemark[4]
                                              &  0.2825(20)\footnotemark[5]
                                                         &   4.233(22)\footnotemark[6]
                                                                      &  10.308(15)\footnotemark[7] \\
                                &  4.4890(65)\footnotemark[8]
                                              &  0.2757(20)\footnotemark[9]
                                                         &            &  \\
Other results                   &  4.5064(47)\footnotemark[10]
                                          &          &            &  \\
\end{tabular}
\end{ruledtabular}
\qquad \qquad \qquad \qquad
\footnotemark[1]{Ref.~\cite{Der01Br}};
\footnotemark[2]{Ref.~\cite{SapChe05}};
\footnotemark[3]{Ref.~\cite{FlaGin05}};
\footnotemark[4]{Ref.~\cite{YouHilSib94}};
\footnotemark[5]{Ref.~\cite{ShaMonKhl79}(as re-evaluated in Ref.~\cite{VasSavSaf02})};
\footnotemark[6]{Ref.~\cite{BouGuePot84}};
\footnotemark[7]{Ref.~\cite{BenRobWie99}};
\footnotemark[8]{Ref.~\cite{RafTanLiv99}};
\footnotemark[9]{Ref.~\cite{VasSavSaf02}};
\footnotemark[10]{Ref.~\cite{DerPor02}};
\end{table*}

Now we proceed to evaluating the PNC amplitude~(\ref{Eq:EPNC}) by
directly summing over the intermediate $nP_{1/2}$ ($n=6-9$)
states~\cite{BluJohSap90}. These states contribute 99\% to the final
$E_{\rm PNC}$ value. In
Table~\ref{Tab:Hw_E1} we present the matrix elements of the electric
dipole operator and $H_W$, as well as the energy differences.
Contributions from the $6-9P_{1/2}$ states to $E_{\rm PNC}$ are also
listed. The matrix elements were computed in the CCSDvT
approximation with dressing. We also used the scaling procedure.
Energy differences were based on the experimental energies with the
Breit and QED corrections removed. For the $6S_{1/2}$, $6P_{1/2}$,
$7S_{1/2}$, and $7P_{1/2}$ states we used the following
``corrected'' energies
\begin{equation}
E_{\rm corr} = E_{\rm exp} - \delta E_{\rm Breit} - \delta E_{\rm QED}.
\label{Eq:Ecorr}
\end{equation}
The contribution of the higher-energy intermediate $8P_{1/2}$ and $9P_{1/2}$ states to the PNC amplitude is
suppressed; for these states we used the full experimental energies.

\begin{table*}[h]
\caption{ Contribution to $E_{\rm PNC}$ from intermediate states
6--9\,$P_{1/2}$. Dipole matrix elements are of the form ${\langle
nLJ;m_J=1/2 \,|D_z| \, n'L'J';m_{J'} =1/2 \rangle }$.}
\label{Tab:Hw_E1}
\begin{ruledtabular}
\begin{tabular}{lcccc}
  & \multicolumn{4}{c}{$6S_{1/2}$ perturbed} \\
$n$
  & \multicolumn{1}{c}{$\langle 7S_{1/2} \, |D| \, nP_{1/2}\rangle$}
  & \multicolumn{1}{c}{$\langle nP_{1/2} \, |H_W| \, 6S_{1/2} \rangle$}
  & \multicolumn{1}{c}{$ E_{6S_{1/2}} - E_{nP_{1/2}}$}
  & \multicolumn{1}{c}{Contribution} \\
  & \multicolumn{1}{c}{a.u.} & \multicolumn{1}{c}{$10^{-11} i (-Q_W/N)$ a.u.}
  & \multicolumn{1}{c}{a.u.} & \multicolumn{1}{c}{$10^{-11} i (-Q_W/N)$ a.u.} \\
\hline
\smallskip
 6    &  1.7327  &  0.05575  &  -0.050949  & -1.8962 \\
\smallskip
 7    &  4.2071  & -0.03169  &  -0.099227  &  1.3435 \\
\smallskip
 8    &  0.3769  & -0.02118  &  -0.117208  &  0.0681 \\
\bigskip
 9    &  0.1423  & -0.01605  &  -0.125993  &  0.0181 \\

  & \multicolumn{4}{c}{$7S_{1/2}$ perturbed} \\
$n$
  & \multicolumn{1}{c}{$\langle nP_{1/2} \, |D| \, 6S_{1/2}\rangle$}
  & \multicolumn{1}{c}{$\langle 7S_{1/2} \, |H_W| \, nP_{1/2} \rangle$}
  & \multicolumn{1}{c}{$ E_{7S_{1/2}} - E_{nP_{1/2}}$}
  & \multicolumn{1}{c}{Contribution} \\
  & \multicolumn{1}{c}{a.u.} & \multicolumn{1}{c}{$10^{-11} i (-Q_W/N)$ a.u.}
  & \multicolumn{1}{c}{a.u.} & \multicolumn{1}{c}{$10^{-11} i (-Q_W/N)$ a.u.} \\
\hline
\smallskip
 6    & -1.8402  & -0.02697  &   0.033573  &  1.4783 \\
\smallskip
 7    &  0.1134  &  0.01525  &  -0.014705  & -0.1176 \\
\smallskip
 8    &  0.0305  &  0.01024  &  -0.032686  & -0.0096 \\
\bigskip
 9    &  0.0128  &  0.00776  &  -0.041471  & -0.0024 \\

Total &          &           &             &  0.8823
\end{tabular}
\end{ruledtabular}
\end{table*}

The results for the PNC amplitude are presented in
Table~\ref{Tab:PNC}. Each subsequent line in the upper
panel of the table corresponds to an increasingly more complex
approximation. We start from the SD approximation. Inclusion  of
non-linear CC terms (``CC'' entry) modifies the SD result by almost
2\%. At the next step we incorporate valence triples. Due to the
importance of these terms we present a detailed breakdown of the
associated effects. We distinguish between indirect and direct
contributions from the valence triples. Indirect effects of triples
come from modifying energies and single and double excitations
through the Schr\"{o}dinger equation. In Table~\ref{Tab:PNC} the
relevant values are marked with ``no vT in MEs''. The direct
contribution arises from explicit presence of valence triples in
expressions for the MEs.
We also list the results obtained without scaling [``vT (no vT in MEs;
pure)''] and including it [``vT (no vT in MEs; scaled)'']. We find
that the PNC amplitude is insensitive to scaling. Note that a
similar conclusion was drawn in Ref.~\cite{DzuFlaGin02}. As the next
step we replace the calculated energies in the denominators of
Eq.~(\ref{Eq:EPNC}) by the experimental energies $E_{\rm corr}$ as
explained above. The resulting entries include the $E_{\rm corr}$
qualifier. We also include line and vertex dressing; the resulting
matrix elements and detailed breakdown of results are listed in
Table~\ref{Tab:Hw_E1}. We also add contributions of intermediate
states above $9P_{1/2}$, including continuum, and contributions from
core excitations. These contributions are denoted as ``$n \geq 10$''
and ``Core contribution'', respectively. Finally,  the lower panel
summarizes well-established non-Coulomb contributions such as the
magnetic interaction between the electrons (Breit), radiative (QED),
and other smaller corrections.

The accuracy of the PNC amplitude was estimated by comparing
theoretical results for energies, dipole matrix elements, and
magnetic hyperfine constants with high-precision experimental data
(see Tables \ref{Tab:Cs_E}--\ref{Tab:E1_ME}).
We find that the experimental energies are reproduced with an accuracy of 0.1-0.3\%.
Relevant dipole matrix elements are within the error bars of the
experiments. Finally, since  the hyperfine constants $A$ are accumulated in the
nuclear region, matrix elements of the weak interaction $\langle
nS_{1/2} |H_W| n'P_{1/2} \rangle$ may be tested by forming the
geometric mean $\sqrt{A_{nS_{1/2}} A_{n'P_{1/2}}}$,
Ref.~\cite{DzuFlaGin02}. We find that the standard deviation of
theoretical values from experiment is 0.2\%. As a test of stability
of the final result, we also computed the main term using {\em ab
initio} (i.e., without scaling) CCSDvT matrix elements and energies.
The result, 0.8839, deviates by 0.18\% from our scaled value of
0.8823 in Table~\ref{Tab:PNC}. Based on these tests, we assign an
error of 0.2\% to the main term. Finally, the ``tail'' lumps
contributions of remaining excited $nP_{1/2}$ states (including
continuum) and core-excited states. The tail was computed using a
blend of many-body approximations and we assign a  10\% uncertainty
to this contribution based on the spread of its value in different
approximations. The final result
includes smaller non-Coulomb corrections and its uncertainty was
estimated by adding individual uncertainties in quadrature.

\begin{figure}[h]
\begin{center}
\includegraphics*[scale=0.5]{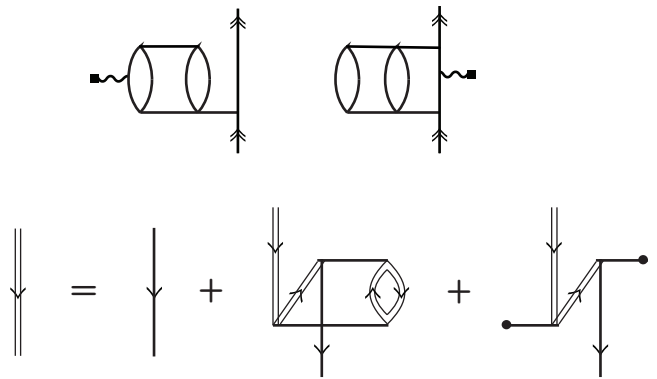}
\caption{ Many-body diagrams responsible for the shift of the PNC
amplitude compared to previous calculations.
 Top row: sample direct contributions of valence triples to matrix elements (wavy capped line)~\protect\cite{PorDer06Na}. Bottom row: iterative equation for line dressing of the hole line in expressions for matrix elements~\cite{DerPor05} (similar equation holds for particle lines; exchange diagrams are not shown).
\label{Fig:diags}}
\end{center}
\end{figure}

Our uncertainty in $E_\mathrm{PNC}$ represents a two-fold
improvement over calculations~\cite{DzuFlaGin02} and a four-fold
improvement over Ref.~\cite{BluJohSap90}. Both calculations report a
value of 0.908 for the total Coulomb-correlated value, larger by
0.9\% than our 0.27\%-accurate result. The reason for the shift in
our more complete calculations is three-fold: (i) direct
contributions of the triple excitations to matrix elements (0.3\%),
(ii) line-dressing of diagrams for matrix elements (0.3\%), and
(iii) consistent removal of Breit and QED corrections from
experimental energies used in the scaling procedure (0.3\%).
Representative diagrams are shown in Fig.~\ref{Fig:diags}.

As discussed, we make a distinction between the indirect and direct
contributions of $T_v$ to matrix elements \cite{PorDer06Na}.
Indirect effects of triples come from modifying energies and single
and double excitations. In the previous
work~\cite{DzuFlaGin02,BluJohSap90} it was approximately accounted
for by a semiempirical scaling of single valence excitation (or
Brueckner orbitals) to the ratio of the theoretical to experimental
correlation energies. Direct $T_v$ contributions to matrix elements,
however, cannot be reproduced by the scaling and, moreover, require
storing triples; due to large-memory requirements this was not done
in Ref.~\cite{DzuFlaGin02,BluJohSap90}. The size of the effect, $-0.0029$,
is given by the difference between entries ``vT (no vT in MEs;
scaled, $E_{\rm corr}$)''  and ``vT (scaled, $E_{\rm corr}$)'' in
Table~\ref{Tab:PNC}. The line-dressing~\cite{DerPor05} was also not
attempted previously. The line-dressing comes from resumming
non-linear contributions to wave functions,
Eq.~(\ref{Eq:PsivOmega}), in expressions for matrix elements. A
structure of the all-order equations for the dressed hole lines is
presented in Fig.~\ref{Fig:diags}. The value of the line-dressing
correction, $-0.0031$, is listed in Table~\ref{Tab:PNC}.
The direct $T_v$ contributions are most
pronounced for the $6S_{1/2}-7P_{1/2}$ dipole amplitude, where they
shift the value by 3\%; their omission leads to a $4 \sigma$
deviation from experiment~\cite{VasSavSaf02}. Similarly, discarding
line-dressing shifts the theoretical values of $A_{6P_{1/2}}$ by
0.8\%.

\begin{table}[h]
\caption{Contributions to $E_{\rm PNC}$  in different
approximations. $E_{\rm PNC}$ and $\Delta$ are in units of $10^{-11} i (-Q_W/N)$, where $N=78$
is the number of neutrons in $^{133}$Cs nucleus.
In the upper panel of the table $\Delta$ is the difference between the results given in
this row and the previous row. In the lower panels $\Delta$ determines the respective
contribution to $E_{\rm PNC}$.}
\label{Tab:PNC}
\begin{ruledtabular}
\begin{tabular}{ldd}
Approximation & \multicolumn{1}{c}{$E_{\rm PNC}$}
              & \multicolumn{1}{c}{$\Delta$} \\
\hline
{\it ``Main'' term}: \\
SD                                        &    0.8952 & \\
\smallskip
CC                                        &    0.8800 &  -0.0152 \\
vT (no vT in MEs; pure)                   &    0.8911 &   0.0111 \\
vT (no vT in MEs; scaled)                 &    0.8915 &   0.0004 \\
vT (no vT in MEs; scaled, $E_{\rm corr}$) &    0.8885 &  -0.0030 \\
\smallskip
vT (scaled, $E_{\rm corr}$)               &    0.8856 &  -0.0029 \\
Line dressing                             &    0.8825 &  -0.0031\\
\smallskip
Vertex dressing                           &    0.8823 &  -0.0002 \\
{\it Final main} ($n$=6--9)                     &    0.8823 & \\
\hline
{\it ``Tail''}: \\
$n \geq 10$                        &           &   0.0195 \\
\smallskip
Core contribution                         &           &  -0.0020 \\
Basis extrapolation                       &           &  -0.00006 \\

\smallskip
Total                                     &    0.8998 &\\
\hline
{\it Complementary corrections}: \\
Breit\footnotemark[1]                     &           & -0.0054 \\
QED\footnotemark[2]                       &           & -0.0024 \\
Neutron skin\footnotemark[3]              &           & -0.0017 \\
$e$--$e$ weak interaction\footnotemark[4] &           &  0.0003\\
\smallskip
{\it Sum of corrections}                  &           & -0.0092\\
Final $E_{\rm PNC}$                       &   0.8906 &\\
\end{tabular}
\end{ruledtabular}
\footnotemark[1]{Ref.~\cite{Der01Br}};
\footnotemark[2]{Ref.~\cite{ShaPacTup05}};
\footnotemark[3]{Ref.~\cite{Der02}};
\footnotemark[4]{Ref.~\cite{MilSusTer02,MilSusTer03a}}.
\end{table}

\section{Weak nuclear charge and implications for particle physics}
\label{sec_Qw}


With the computed $E_\mathrm{PNC}$  we proceed to extracting the electroweak observable.
The experiment~\cite{WooBenCho97} determined the
ratio of $E_\mathrm{PNC}/\beta=1.5935(56) \, \mathrm{mV/cm}$. The most accurate $\beta$ comes from a combined
determination~\cite{BenWie99,DzuFlaGin02}, $\beta=-26.957(51) a_B^3$.  As a result we arrive at the nuclear weak charge
\begin{equation}
 Q_W(^{133}\mathrm{Cs}) = -73.16(29)_\mathrm{exp}(20)_\mathrm{th} \, , \label{Eq:QwResult}
\end{equation}
where the first uncertainty is experimental and the second uncertainty is theoretical.
Taking a weighted average, $\beta=-26.99(50) a_B^3$, of two determinations~\cite{BenWie99,VasSavSaf02}
results in $Q_W(^{133}\mathrm{Cs}) = -73.25(29)_\mathrm{exp}(20)_\mathrm{th}$. Both values are in a perfect agreement
with the prediction of the standard model, $Q_W^\mathrm{SM} =-73.16(3)$ of Ref.~\cite{Ams08}.

While our  result is consistent with the SM, it plays a unique and at the same time complementary role to
high-energy physics experiments.
Our result (i) confirms energy-dependence (or running) of the electroweak interaction and (ii) places constraints
on a variety of new physics scenarios beyond the standard model.

\begin{figure}[h]
\begin{center}
\includegraphics*[scale=0.75]{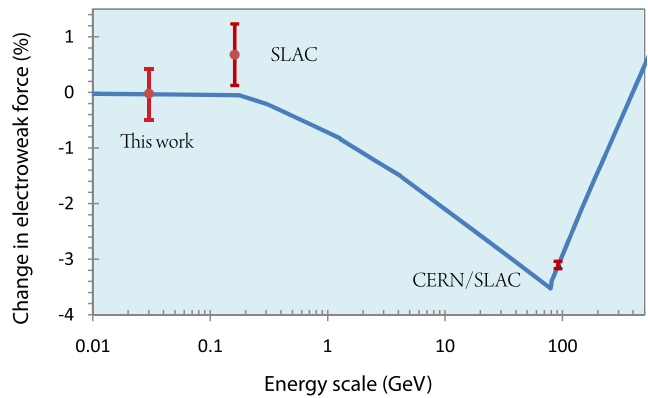}
\caption{(Color online) Running of the electroweak force.
The strength of the electroweak coupling varies depending on the energy scale
probed by an experiment. The plot shows the amount of variation relative to
the strength at zero energies. The solid line is the prediction~\cite{ErlRam05} of the SM. High-energy experiments
at CERN and SLAC have measured the strength of electroweak force at 91 GeV with an accuracy of $\sim 0.1$\%.
In 2005, a SLAC electron-scattering experiment~\cite{AntArnArr05}  had determined the strength
at 0.2 GeV with an accuracy of about 0.5\%.
Our analysis of atomic parity violation probes the least energetic (30 MeV) electroweak interactions measured so far and the result is in perfect agreement with the SM. Overall, the predicted running of the electroweak force is confirmed over an energy range spanning 4 orders of magnitude.
\label{Fig:run}}
\end{center}
\end{figure}

In physics, the vacuum is never still. Each particle carries a cloud of continuously
sprouting virtual particle-antiparticle pairs. The strength of the mutual interaction between
two particles becomes dependent on their relative collision energy: at higher energies,
the collision partners tend to penetrate deeper inside the shielding clouds.
According to the SM, the interaction strength at low energies differs by about 3\% from its value at 100 GeV,
see Fig.~\ref{Fig:run}.
For low energies, where the shielding clouds are
penetrated the least, previous analyses~\cite{ShaPacTup05,DerPor07} were consistent with no running.
Here we improve the accuracy of probing these least energetic electroweak interactions.


Compared to conventional particle-physics experiments, our result provides a reference point for the least energetic electroweak interactions.
With our weak charge, we find the effective interaction strength,  $\sin^2 \theta_W^\mathrm{eff}(E \rightarrow 0) =0.2382(11)$. The result is in agreement with the SM value~\cite{CzaMar00} of 0.2381(6). While an earlier evidence for running of $\sin^2 \theta_W$ has been obtained in the parity violating electron scattering experiment at SLAC~\cite{AntArnArr05}, the prediction of the SM was outside their experimental error bars. Our work provides a higher-confidence confirmation of the predicted running of the electroweak coupling at low energies.

Notice that the relevant momentum transfer for $^{133}$Cs atom is just $\sim\! 30$ MeV, but the exquisite accuracy of
the interpretation probes minute contributions of the sea of virtual (including so-far undiscovered) particles
at a much higher mass scale. The new physics brought by the virtual sea is phenomenologically
described by weak isospin-conserving $S$ and isospin-breaking $T$ parameters~\cite{Ros02}: $Q_W -Q_W^\mathrm{SM} = -0.800\, S - 0.007 T$.
At the $1\,\sigma$-level, our result implies $|S|<0.45$. The parameter $S$ is important, for example, in indirectly constraining the mass of the Higgs particle~\cite{Ros02}.
Similarly, the extra $Z$ boson, $Z'_\chi$, discussed in Ref.~\cite{PorBelDer09}, would lead to a deviation~\cite{MarRos90}
$Q_W -Q_W^\mathrm{SM} \approx 84 (M_W/M_{Z'_\chi})^2$. We find $M_{Z'_\chi} > 1.4 \, \mathrm{TeV}/c^2$, improving
the present lower bound on the $Z'$ mass from direct collider searches~\cite{AalAbuAde07}.

\section{Acknowledgments}
\label{sec_Ac}
We thank O.~Sushkov, M.~Kozlov, J.~Erler, W.~Marciano, and M.~Ramsey-Mussolf for
discussions.
This work was initiated with support from the NIST precision measurement grant program and
supported in part by the NSF.



\end{document}